\documentclass{iaus}
\usepackage{graphicx}

\title[Chemical clocks for early-type galaxies] 
{Chemical clocks for early-type galaxies in clusters}

\author[Carretero, Vazdeks \& Beckman]   
{Conrado Carretero$^1$, Alexandre Vazdekis$^1$ \and John E. Beckman$^{1,2}$}

\affiliation{$^1$Instituto de Astrof\'{\i}sica de Canarias. 38200 La Laguna, Tenerife, Spain. Email: cch@iac.es\\[\affilskip]
$^2$Consejo Superior de Investigaciones Cient\'{\i}ficas, Spain.}

\pubyear{2007}
\volume{241}  
\pagerange{119--126}
\date{?? and in revised form ??}
\jname{Proceedings Title IAU Symposium}
\editors{A.C. Editor, B.D. Editor \& C.E. Editor, eds.}
\begin{document}

\maketitle

\begin{abstract}
We present a detailed stellar population analysis of 27 massive elliptical galaxies within 4 very rich clusters at redshift $z$$\sim$0.2. We obtained accurate estimates of the mean luminosity-weighted ages and relative abundances of CN, Mg and Fe as functions of the galaxy velocity dispersion, $\sigma$. Our results are compatible with a scenario in which the stellar populations of massive elliptical galaxies, independently of their environment and mass, had formation timescales shorter than $\sim$1~Gyr. This result implies that massive elliptical galaxies have evolved passively since, at least, as long ago as $z$$\sim$2. For a given galaxy mass the duration of star formation is shorter in those galaxies belonging to more dense environments. Finally, we show that the abundance ratios [CN/Fe] and [Mg/Fe] are the key ``chemical clocks'' to infer the star formation history timescales in ellipticals. In particular, [Mg/Fe] provides an upper limit for those formation timescales, while [CN/Fe] apperars to be the most suitable parameter to resolve them in elliptical galaxies with $\sigma$$<$300~km~s$^{-1}$.
\end{abstract}

\firstsection 
\section{Introduction}

Stellar Population analysis has recently shown increasing power in constraining galaxy formation scenarios. They offer a fossil record of the star formation and chemical evolution of galaxies, most clearly in elliptical galaxies, thought to have formed by the merging of discs, so stellar population studies provide very strong constraints on these galaxy formation scenarios. For example, it is particularly hard to reconcile the hierarchical models with the result that massive galaxies show significantly older mean luminosity weighted ages than their smaller counterparts (Kauffmann et al. 2003).

A key point is to see how abundance ratios in early-type galaxies vary with the cluster masses. In particular, overabundances of [Mg/Fe] compared with the solar ratio have been found in massive elliptical galaxies (e.g. Vazdekis et al. 1997). In this work, we concentrate on the stellar population analysis of 27 elliptical galaxies distributed in 4 rich, intermediate-redshift ($z$$\sim$0.2) Abell clusters.

\section{Observations and Data}

We have studied a set of 27 elliptical galaxies belonging to 4 very rich  Abell clusters: A115, A655, A963 and A2111. The selected galaxies are the brightest galaxies in their respective clusters (15.1$<$$m_V$$<$17.2) and have been already morphologically analyzed in the literature. The velocity dispersion covers the range 220~km~s$^{-1}$$<$$\sigma$$<$450~km~s$^{-1}$.

Multi-slit spectroscopy was carried out with the DOLORES spectrograph on the 3.5m TNG Telescope (La Palma). We designed MOS masks with 1.1$''$ width slits and an instrumental resolution of 8 \AA\ (FWHM).

\section{Analysis}

To derive mean luminosity-weighted ages and metallicities, we compared selected absorption line strengths with those predicted by the model of Vazdekis et al. (2006, in preparation). This model provides flux-calibrated spectra in the wavelength range $\lambda\lambda$~3500-7500 \AA, at a resolution of 2~\AA\ (FWHM) for single-burst stellar populations.

\begin{figure*}
\resizebox{\hsize}{!}{\includegraphics{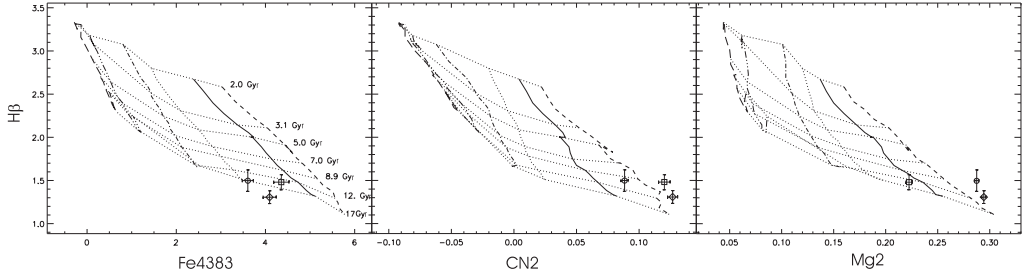}}
\caption{Examples of the method used to derive ages and metallicities. The plot shows the age index H$\beta$ versus the metallicity indicators Fe4383, CN2 and Mg$_2$ for the grid of models of Vazdekis et al. (2006, in preparation). Horizontal dotted lines represent models of constant age whose ages are labelled in the plot. Vertical dashed, solid, dotted, dashed-dotted, dashed-dotted-dotted and long-dashed lines represent contours of constant [M/H]~=+0.2, 0.0, -0.4, -0.7, -1.2 and -1.7 dex, respectively. Diamonds, open circles and squares indicate the values for a representative galaxy of A115, A655 and A963, respectively, with a common $\sigma$=310~km~s$^{-1}$. Note that all synthetic model spectra were broadened to the instrumental resolution of the observed spectra and the velocity dispersion of these particular galaxies ($\sigma_{\rm obs}$=350~km~s$^{-1}$).}
\label{modelgrid}
\end{figure*}

Once the model spectra are transformed to yield the intrumental conditions of resolution and dispersion of the observed spectrum, we measure pairs of indices in both sets of data (observed and synthetic). Figure \ref{modelgrid} illustrates this method.

\section{Results}

\begin{figure*}
\includegraphics[width=83mm]{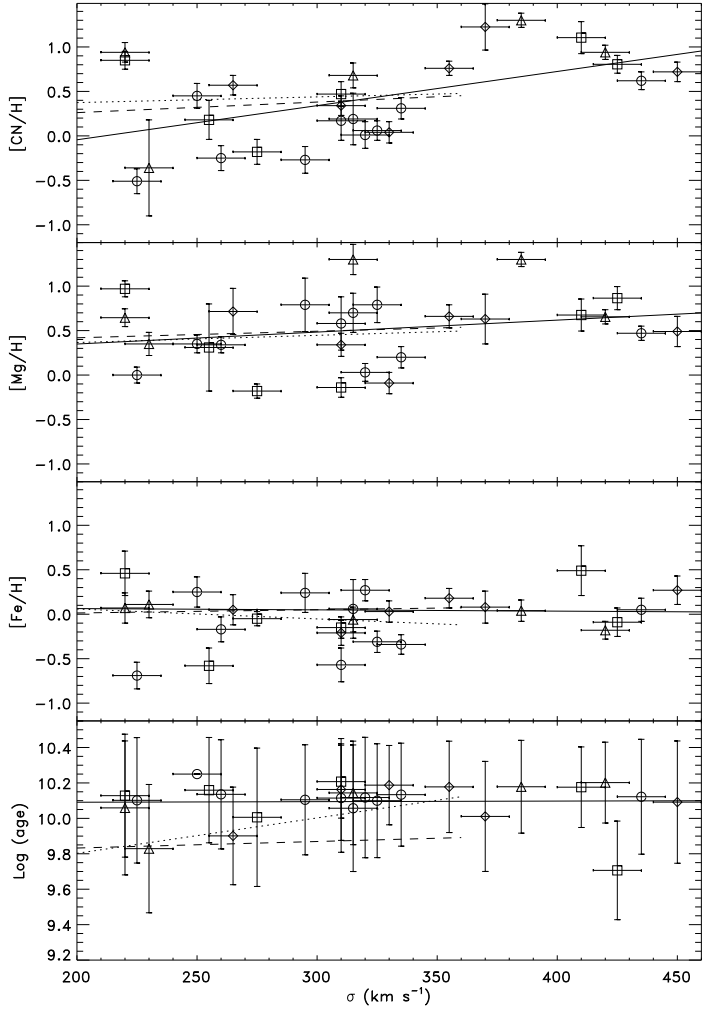}\centering
\vspace{1cm}
\caption{Relations between the relative abundances [CN/H], [Mg/H], [Fe/H], and mean luminosity-weighted ages and the velocity dispersion of the galaxies, $\sigma$. Diamonds: A115 galaxies. Open circles: A655 galaxies. Open squares: A963 galaxies. Open triangles: A2111 galaxies. The solid line corresponds to the error-weighted linear fitting to our data. Dotted line and dashed line corresponds to the relations obtained for Virgo cluster galaxies and Coma cluster galaxies, respectively. It is noteworthy that the slope of the relations increases with the environment density for [CN/H]-$\sigma$, while it is almost constant in the case of [Mg/H]-$\sigma$ and [Fe/H]-$\sigma$.}
\label{abundancessigma}
\end{figure*}

Figure \ref{abundancessigma} presents the relations [CN/H]-$\sigma$, [Mg/H]-$\sigma$, [Fe/H]-$\sigma$ and Log(age)-$\sigma$ for our full sample. We have also plotted on Figure \ref{abundancessigma} the results obtained by S\'anchez-Bl\'azquez et al. (2006), who analyzed 98 early-type galaxies of the Virgo and Coma clusters.

\begin{figure*}
\includegraphics[width=83mm]{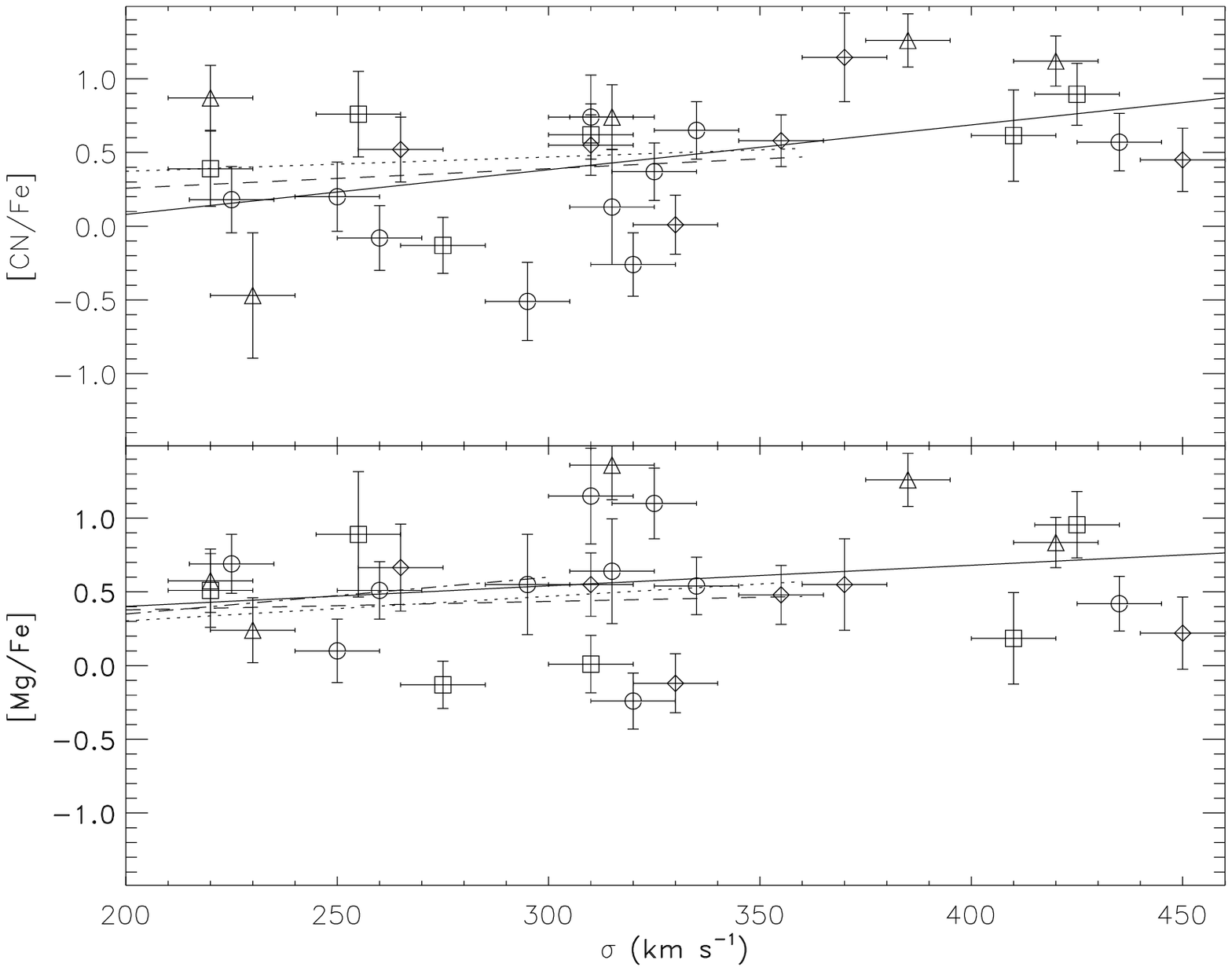}\centering
\vspace{1cm}
\caption{Abundance ratios [CN/Fe] and [Mg/Fe] versus the velocity dispersion of the galaxies. Symbols are the same than in previous Figure. Dashed-dotted line in bottom panel corresponds to the [Mg/Fe]-$\sigma$ relation for Virgo cluster galaxies, by Vazdekis et al. (2004). Note that the slope of the [CN/Fe]-$\sigma$ relation increases with the density of the environment, but the [Mg/Fe]-$\sigma$ remains almost constant.}
\label{abundanceratios}
\end{figure*}

Figure \ref{abundanceratios} presents the relations of the abundance ratios [CN/Fe] and [Mg/Fe] with the velocity dispersion. We observe positive trends with $\sigma$ for both ratios.

\section{Discussion}

First, when inspecting Figures \ref{abundancessigma} and \ref{abundanceratios} we observe that, for a given $\sigma$, the relative abundance [CN/H] increases with the cluster X-ray luminosity, whereas [Mg/H] and [Fe/H] do not. The same happens with the abundance ratios: while [CN/Fe] increases as a function of the environment density, the [Mg/Fe] ratio keeps constant and {\em positive}.

If we consider the formation timescales for Mg and Fe we can estimate that, whatever the environment is, all stellar populations in ellipticals should be already formed in less than $\sim$1~Gyr, since this is the time needed by SNe~Ia to pollute the ISM with iron-peak elements. But, if we consider species with less separated formation timescales, such as Fe and CN, we do find differences in the ratio [CN/Fe] as a function of the environment, for a given $\sigma$. That suggests a different formation timescale for the stars in ellipticals related to the environment properties, in the sense that the more dense the environment is the shorter the formation timescales of the stars are (with an upper limit of $\sim$1~Gyr). This is in agreeement with a number of recent results (e.g. Carretero et al. 2004, Thomas et al. 2005, Bernardi et al. 2006).

Second, it is noteworthy that the slope of the relation [CN/Fe]-$\sigma$ increases with the cluster X-ray luminosity while the slope of the [Mg/Fe]-$\sigma$ is constant with the cluster mass. Because of that, we can only find environment-related [CN/Fe] differences in those galaxies with $\sigma$$<$300~km~s$^{-1}$ whereas, for more massive galaxies, the [CN/Fe] values are almost equal independently of the cluster properties. This means that the ratio [CN/Fe] in most massive galaxies is less sensitive to the environment than in intermediate- and low-mass galaxies, suggesting that star formation histories are less environment dependent as the galaxy mass increases. Therefore, [CN/Fe] appears to be an appropriate ``chemical clock'' for ellipticals with $\sigma$$<$300~km~s$^{-1}$ but, when we move to very massive galaxies, the formation timescales differ so gently from one environment to another that this ``clock'' turns out to be insufficiently sensitive. In order to disentangle the different formation timescales between very massive galaxies in different environments, it is neccesary to find out a more accurate indicator.

Our results suggest an overall picture in which massive elliptical galaxies in very rich clusters are old systems, with very short star formation histories, and which have been passively evolving since, at least, $z$$\sim$2. This is in agreement with a number recent results (e.g. Lab\'e et al. 2005, van Dokkum et al. 2006). In comparison with less dense environments, the stars in elliptical galaxies of rich clusters must have formed at slightly earlier epochs and on a slightly shorter timescales.

\begin{acknowledgments}
This work has been supported by the Spanish Ministry of Education and Science grants AYA2004-03059 and AYA2004-08251-C02-01.
\end{acknowledgments}


\begin{thebibliography}{}

\bibitem[\protect\citeauthoryear{{Bernardi}, {Nichol}, {Sheth}, {Miller} \&
  {Brinkmann}}{{Bernardi} et~al.}{2006}]{bernardi06}
{Bernardi} M.,  {Nichol} R.~C.,  {Sheth} R.~K.,  {Miller} C.~J.,    {Brinkmann}
  J.,  2006, AJ, 131, 1288

\bibitem[\protect\citeauthoryear{{Carretero}, {Vazdekis}, {Beckman},
  {S{\'a}nchez-Bl{\'a}zquez} \& {Gorgas}}{{Carretero}
  et~al.}{2004}]{carretero04}
{Carretero} C.,  {Vazdekis} A.,  {Beckman} J.~E.,  {S{\'a}nchez-Bl{\'a}zquez}
  P.,    {Gorgas} J.,  2004, ApJ, 609, L45

\bibitem[\protect\citeauthoryear{{Kauffmann}, {et al.}}{{Kauffmann} et~al.}{2003}]{kauffmann03}
{Kauffmann} G.,  {et} al.,  2003, MNRAS, 341, 54

\bibitem[\protect\citeauthoryear{{Labb{\'e}}, {et al.}}{{Labb{\'e}}
  et~al.}{2005}]{labee05}
{Labb{\'e}} I.,  {et} al.,  2005, ApJ, 624, L81

\bibitem[\protect\citeauthoryear{{S{\'a}nchez-Bl{\'a}zquez}, {Gorgas},
  {Cardiel} \& {Gonz{\'a}lez}}{{S{\'a}nchez-Bl{\'a}zquez}
  et~al.}{2006}]{sanchezblazquezetal06}
{S{\'a}nchez-Bl{\'a}zquez} P.,  {Gorgas} J.,  {Cardiel} N.,    {Gonz{\'a}lez}
  J.~J.,  2006, A\&A, 457, 809

\bibitem[\protect\citeauthoryear{{Thomas}, {Maraston}, {Bender} \& {Mendes de
  Oliveira}}{{Thomas} et~al.}{2005}]{thomas05}
{Thomas} D.,  {Maraston} C.,  {Bender} R.,    {Mendes de Oliveira} C.,  2005,
  ApJ, 621, 673

\bibitem[\protect\citeauthoryear{{van Dokkum} \& {van der Marel}}{{van Dokkum}
  \& {van der Marel}}{2006}]{vandokkum06}
{van Dokkum} P.~G.,  {van der Marel} R.~P.,  2006, astro-ph/0603063

\bibitem[\protect\citeauthoryear{{Vazdekis}, {Peletier}, {Beckman} \&
  {Casuso}}{{Vazdekis} et~al.}{1997}]{vazdekisetal97}
{Vazdekis} A.,  {Peletier} R.~F.,  {Beckman} J.~E.,    {Casuso} E.,  1997,
  ApJS, 111, 203

\bibitem[\protect\citeauthoryear{{Vazdekis}, {Trujillo} \& {Yamada}}{{Vazdekis}
  et~al.}{2004}]{vazdekistrujillo04}
{Vazdekis} A.,  {Trujillo} I.,    {Yamada} Y.,  2004, ApJ, 601, L33




\end{thebibliography}
\end{document}